%% file: preondarev2.tex
\documentstyle[preprint,prl,aps]{revtex}

\newcommand{\be}{\begin{equation}}
\newcommand{\ee}{\end{equation}}
\newcommand{\ben}{\begin{eqnarray}}
\newcommand{\een}{\end{eqnarray}}
\newcommand{\un}{\underline}

\newcommand{\ra}{\rangle}
\newcommand{\la}{\langle}
\newcommand{\ram}{\rangle_{\mu}}
\newcommand{\lam}{{}_{\mu}\langle}
\newcommand{\ov}{\overline}
\newcommand{\kn}{| n \rangle}
\newcommand{\bn}{ \langle n |}

\newcommand{\kvi}{|i {\ram}}

\newcommand{\til}{\tilde}
\newcommand{\iii}{\'{\i}}
\newcommand{\sep}{\;\;\;\;\;\;;\;\;\;\;\;\;}
\newcommand{\spa}{\;;\;}

\newcommand{\sumn}{\sum_{n=1}^N}
\newcommand{\sumk}{\sum_{j=1}^k}

\newcommand{\sumkk}{\sum_{n=1}^{k+1}}

\newcommand{\pp}{p^{\frac{1}{2}}}
\newcommand{\ppkk}{p^{\frac{1}{2}(k+1)}}
\newcommand{\ppk}{p^{\frac{1}{2}(k)}}

\newcommand{\lamb}{\lambda}
\newcommand{\sii}{\sigma_i}

\newcommand{\kaln}{|\alpha_{l_n}\ram}
\newcommand{\baln}{\lam \alpha_{l_n}|}

\newcommand{\op}{\hat{P}}
\newcommand{\iov}{\hat{I}_\mu}

\newcommand{\OPK}{\hat{P}_{V_{k}}}
\newcommand{\OPKK}{\hat{P}_{V_{k+1}}}

\newcommand{\OPKKT}{\hat{P}_{V_{k+1}}^\dagger}

\newcommand{\kp}{|\pp \rangle}

\newcommand{\AT}{\hat{A}_\mu^\dagger}
\newcommand{\AO}{\hat{A}_\mu}

\newcommand{\FOK}{\hat{F}_k}

\newcommand{\R}{{\cal{R}}}
\newcommand{\RN}{{\R}^N}
\newcommand{\B}{{\cal{D}}}
\newcommand{\BM}{{\cal{D}}^M(\mu)}

\newcommand{\spann}{{\mbox{span}}}

\newcommand{\kg}{ | g \ram}
\newcommand{\bg}{ \lam  g |}
\newcommand{\kfe}{ | f^p \ram}

\newcommand{\kfo}{ | f^o \ram}
\newcommand{\kfot}{ | {\til{f}}^o \ram}
\newcommand{\bfot}{ \lam {\til{f}}^o |}
\newcommand{\kfn}{ | f_n \rangle_{\mu}}
\newcommand{\bfn}{\lam f_n |}

\newcommand{\kpsik}{|{\psi}_{k+1} \ram}
\newcommand{\bpsik}{\lam {\psi}_{k+1} |}

\newcommand{\kpsiktt}{| \til{\til{\psi}}_{k+1}\ram}
\newcommand{\bpsiktt}{\lam \til{\til{\psi}}_{k+1}|}

\newcommand{\balkkkts}{\lam \til{\alpha}_{l_{k+1}}^{k+1} |}

\newcommand{\kalnkts}{|\til{\alpha}_{l_n}^k \ram}
\newcommand{\balnkts}{\lam \til{\alpha}_{l_n}^k|}
\newcommand{\kalnkkts}{|\til{\alpha}_{l_n}^{k+1} \ram}
\newcommand{\balnkkts}{\lam \til{\alpha}_{l_n}^{k+1}|}

\newcommand{\sumnj}{\sum_{n=1 \atop n \ne j}^k}

\newcommand{\Nj}{k/j}
\newcommand{\vn}{V_{k}}
\newcommand{\vnj}{V_{k/\alpha_{l_j}}}

\newcommand{\pvn}{\hat{P}_{V_k}}
\newcommand{\pvnj}{\hat{P}_{\vnj}}

\newcommand{\altt}{\til{\alpha}}
\newcommand{\kalj}{{|\alpha}_{l_j}\ram}

\newcommand{\kaltjN}{|\altt_{l_j}^k \ram}
\newcommand{\baltjN}{\lam{\altt}_{l_j}^k |}

\newcommand{\kaltnN}{|\til{\alpha}_{l_n}^k \ram}

\newcommand{\kaltnNj}{|\til{\alpha}_{l_n}^{\Nj} \ram}
\newcommand{\baltnNj}{\lam \til{\alpha}_{l_n}^{\Nj}|}

\begin{document}

\title{Constructive approximations of 
the $q=1/2$ MaxEnt distribution from redundant and 
noisy data} 

\author{L.Rebollo-Neira} 
\address{NCRG, Aston University\\ 
Birmingham B4 7ET, United Kingdom}

\author{A. Plastino} 
\address{Instituto de  F\iii sica La Plata (IFLP) \\
Universidad Nacional de La Plata and CONICET\thanks
{Argentina's National Research Council}
\\ C.C. 727, 1900 La Plata, Argentina}
\draft

\maketitle

\begin{abstract}

The problem of constructing the $q=1/2$ non-extensive maximum
entropy distributions from redundant and noisy data is
considered. A strategy is proposed, which evolves 
through the following steps: 
i)independent constraints are first pre-selected  by
recourse to a data-independent technique to be discussed 
here.
ii)the data are a
posteriori used to determine the parameters of the distribution
by a previously introduced forward approach. iii) A  backward  
approach is proposed for 
reducing the parameters of such distribution. 
The previously introduced forward approach
is generalised here in order to make it
suitable for dealing with very noisy data.

\end{abstract}
\newpage

\section{Introduction}
Among the generalised non-extensive MaxEnt 
distributions, which are defined in terms of a parameter
$q$ \cite{t1,t2,pla1} the corresponding to
the value $q=1/2$ has played  a 
  particular  role in 
 diverse contexts 
\cite{bo,pa1,pa2,pre1,pre2,nu}.

In this paper we focus on developing strategies for 
constructing the $q=1/2$ distribution which is involved 
in a very special type of inverse problem: 
the problem of constructing such a distribution 
on the basis of redundant and noisy data 
(by noise we mean errors resulting from the 
random process associated to the experimental 
measurement procedure). 

It is appropriate to start by discussing why 
we shall restrict consideration to the particular 
value $q=1/2$. 

The problem of determining a $p^q$ probability
distribution maximising the entropy
$$S_q = \frac{\sumn p_n^q - \sumn p_n}{1-q}$$
with constraints
\ben
f^o_i&=&\sum_{n=1}^N p_n^{q}f_{i,n}
\;\;\;\;\; ;\;\;\;\;\;i=1,\ldots,M\nonumber\\
 1&=& \sum_{n=1}^N p_n^{q} \nonumber
\een
has been shown in \cite{pa2} to be numerically 
equivalent to determining the probability 
distribution $\til{p}$
minimising 
$$||\til p||_{\frac{1}{q}}^{\frac{1}{q}}= 
\sumn \til{p}_n^{1/q}$$
with constrains
\ben
f^o_i&=&\sum_{n=1}^N \til{p}_n f_{i,n}
\;\;\;\;\; ;\;\;\;\;\;i=1,\ldots,M.\nonumber\\
1 &=&\sum_{n=1}^N \til{p}_n.\nonumber
\een
Since $\til{p}_n >0$ it is true that 
$||\til p||_{\frac{1}{q}}$ is the
${\frac{1}{q}}$-norm of $\til p$. Thus, 
the problem of choosing the parameter
$q$ is equivalent to deciding which norm one
wants to minimise as preserving the $1$-norm
of the distribution.
In order to analyse the situation further
let us joint all constrains together
by defining a $(M+1) \times N$ matrix $\widetilde{A}$
of elements 
$\widetilde{A}_{i,n}=f_{i,n};\spa i=1,\ldots,M; \spa
n=1,\ldots,N$ and $ \widetilde{A}_{M+1,n}=1\spa 
n=1,\ldots,N$. Hence, the  constraints   are
expressed in the form
$$f^o=\widetilde{A}{p},$$
where $f^o$ is a vector of $(M+1)$ components
$f^o_1,\ldots,f^o_M,1$. 
It is well know from linear algebra that the general
solution to this under-determined linear system
can be expressed as
$$\til{p} = \widetilde{A'}^{-1} f^o + p'$$
where $\widetilde{A'}^{-1}$ is the pseudo 
inverse of $\widetilde{A}$, 
and $p'$ a vector in the null space of 
matrix $\widetilde{A}$.
Consequently, the problem of deciding
on the $q$-parameter is tantaumont 
to just choosing a
vector $p'$ in the null space of $\widetilde{A}$.
In particular, the choice $q=1/2$ (which as
already discussed is equivalent to minimising
the 2-norm of the $\til{p}$ distribution)
implies to set $p'=0$. This follows from the
fact that, since vector $\widetilde{A}^{-1}f^o$ 
and vector $p'$ are orthogonal with each other one has
$$||\til{p}||^2_2= ||\widetilde{A}^{-1} f^o||^2_2 + 
||p'||^2_2.$$
Hence, by setting $p'=0$ the solution of
minimum 2-norm is obtained.
For a number of reasons, that we spell out below,
we believe that this leads to the most suitable 
choice for the parameter $q$ in relation to our problem.
Indeed,
\begin{itemize}
\item 
The under-determined problem we have to solve is
of the following especial nature: We have
less independent equation than unknowns,
but there is a large number of redundant  equations 
and a number of irrelevant ones \cite{pre1}. 
If the data were noiseless,
the role of such equations would be simply
to verify the ability of the distribution
to make correct predictions.  Since the data are
noisy we use all the equations with the
purpose of reducing the effect of the noise,
but not as independent constraints (in most cases the
number of Lagrange multiplies is much less that
the actual number of available constraints).
Our task is to identify a subset of such independent
constraints. The predictive power of our solution 
is  assessed a posteriori by 
its capability of predicting the denoised data.
\item
The constraints typically represent
measurements obtained as a function of some
variable parameters: Intensity
vs. diffraction angle, magnetisation vs. magnetic field 
etc. \cite{rea,rea2}. 
It is then natural to represent such measurements
as linear functionals on the identical
vector. Each linear functional provides a
projection on the particular parameter
value which is specified by the measurement instrument
state \cite{rea}. It is clear then that 
in the space of the data it is appropriate to 
define a distance through the norm induced by
the inner product.
In our formalism both the space of the data and
the space of the system are assumed to be 
Hilbert spaces. The only $1/q$-norm induced by a Hilbert
space is the one corresponding to $q=1/2$.
\item
As mentioned above, to choose a value of
$q$ other than $q=1/2$ would imply to let the  corresponding 
distribution have a component in
the null space of the transformation generated
by the constraints. In the type of
problem described in the previous item
such a null space is of a `chaotic' nature (in the sense that
arbitrarily small numerical perturbation on any of
the elements of matrix $\widetilde{A}$ would produce and
enormous distortion in the solution). We certainly
wish to avoid this. 
\end{itemize}
Unfortunately, in our context deciding on the 
appropriate  $q$-value of the distribution 
we wish to construct does not solve 
the problem of its optimal construction. 
While it is true that the problem of 
determining the $q=1/2$ distribution
from a fixed set of constraints is a simple 
linear problem \cite{pa1}, the problem becomes 
highly non linear
when this distribution is to be determined optimally from
a subset of constraints which are taken out of a much larger
set of possible ones.

Consider that from  a set of $M$ constraints we want to 
select a subset of $k$ ones and associate a 
parameter (Lagrange multipliers) to each equation. 
Let us indicate as 
$\ppk$ the distribution associated to the 
corresponding $k$ equations. Hence the problems 
we have to face are the following a) the selection of the 
optimal $k$ constraints b) the estimation of the 
corresponding $k$ parameters determining  the 
distribution.  In order to address these problems
 let us  specify the meaning of 
`optimal selection' in our context: 
{\it{we say that a selection is optimal if it yields   
a distribution capable of satisfactorily 
predicting all the available data  involving the
minimum number of parameters}}. Unfortunately 
the search for such an optimal selection is not in general 
possible, as it poses a NP-hard problem, i.e., unreachable in 
polynomial time with classical computers \cite{qc1,qc2}. 
Hence we are forced 
to ascertain suitable suboptimal strategies, which 
also poses an open problem because there is not a  
unique way of constructing suboptimal solutions.

In some recent publications we have introduced a
suboptimal iterative strategy, which is only optimal
at each iteration step \cite{pre1,pre2}. Such an
approach is a forward data dependent approach
for subset selection.
At each iteration
the indices obtained in the previous steps are fixed, and
a new index is chosen in such a way that the distance
between the observed data and the ones predicted by the
physical model is minimised. Since the selection is only 
optimal at each step, the selected set of indices 
is, of course, not optimal in the above specified sense.
Some indices that are relevant at a particular 
step may become much less relevant at the end of 
the process. It is then natural to try and eliminate 
the parameters corresponding to such indices.
Again, the process of reducing 
parameters in an optimal way is in general an 
NP-problem and we need to address it by suboptimal strategies.
Here we propose a strategy for reducing parameters 
that we call backward selection. This new approach  
provides both the criterion for selecting 
the parameters to be deleted  and the technique 
for properly modifying the ones to be retained. 
An approach for selecting independent constraints in the 
absence of data will also be advanced here, with the 
aim of designing a new suboptimal strategy consisting 
of the following steps:

i)Before the experiment is carried out  
we select a subset of  indices corresponding to 
independent constrains.

ii)The forward selection approach proposed in 
\cite{pre2} is then applied for selecting indices,  
from the pre-selected set, in order to construct the 
distribution when the data are available. 

iii)Finally the backward selection approach is
applied in order to reduce further the number of
parameter of the distribution. 
Such backward selection is made possible in  a 
fast an efficient way by means of a backward adaptive 
biorthogonalization technique.

Before advancing the above described new strategy 
we would like to discuss how is possible
to adapt the strategy
of \cite{pre2} so as to  make
it suitable when dealing with very noisy data.
This is achieved by introducing a vectorial space with
inner product defined with respect to a measure depending on
the experimental data, or their corresponding statistics.

The paper is organised as follows:
The generalisation of the previous approach, to turn 
it suitable when dealing with very noisy data, is
introduced in section II.
Section III discusses the criteria for selecting relevant
constraints.
First the selection criterion proposed in \cite{pre1}
is generalised
and a numerical experiment is presented in order to
illustrate the advantage of such a generalisation.
We then discuss a new data independent selection
criterion. In section IV
we introduce a backward procedure for eliminating
constraints and, consequently,
for properly adapting the concomitant parameters of the
distribution. Sections III and IV provide
the foundations of a new strategy that
we illustrate by a numerical example in Section IV.
The conclusions are drawn in section V.

\section{Generalising the previous approach} 
Let us assume that 
we are given 
$M$ pieces of data $f^o_1,f^o_2,\ldots, f^o_i,
\ldots f^o_M$, each of which is the expectation value
of a random variable that takes values 
$f_{i,n} \;; \; n=1,\ldots, N$ according to the $q=1/2$ 
probability distribution 
$p_n^{\frac{1}{2}} \; ;\; n=1,\ldots, N$ 
\cite{pre1,pre2} i.e., 
\be
f^o_i=\sum_{n=1}^N p_n^{\frac{1}{2}} f_{i,n} 
\;\;\;\;\; ;\;\;\;\;\;i=1,\ldots,M.
\label{sy}
\ee
The data  $f^o_1,f^o_2,\ldots, f^o_i, \ldots f^o_M$  
will be represented as components of a vector $|f^o\ram$
in a vector space, say $\B^M$.   A central aim   
of this contribution is to allow for the possibility 
of assigning a different weight to each 
data. Accordingly, the inner product in 
$\B^M$, that we indicate as $\lam . | . \ram$, is 
defined with respect to a measure $\mu(m)$ 
as follows: For every $f$ and $g$ in $\B^M$
\be
\lam f | g \ram = \sum_{i=1}^M \ov{f}_i \, g_i\,\, \mu_i
\ee
where $\ov{f}_i$ indicates the complex conjugate of 
$f$. In the present situation we deal with
real vectors, thereby, $\ov{f}_i\equiv f_i$.
The data space, 
with the corresponding associated measure, will be denoted 
as $\BM$ and the standard orthogonal basis
in $\BM$
will be represented by vectors  
$\kvi\spa i=1,\ldots,M$. The identity operator in 
$\BM$ is thus expressed as:
\be
\iov = \sum_{i=1}^M |i \ram  \,\, \mu_i \,\,\lam i|, 
\ee
with vectors $\kvi ; i=1,\ldots,M$
satisfying the relations 
\be
\mu_i \,\, \lam i |j \ram = \delta_{i,j} \;\;\;{\mbox{(or 0 
if }} \mu_i=0).
\ee
Accordingly, vector $|f^o\ram$ is expressed
\be
|f^o\ram =  \sum_{i=1}^M |i \ram  \, \mu_i \,\lam i|f^o\ram
= \sum_{i=1}^M  \, \mu_i \, f^o_i |i \ram. 
\ee
The measure $\mu$, rendering   a 
weighted distance between two vectors in $\BM$,  
will be chosen in relation to 
the observed data. For example, if the variances of the 
 data are known and we denote by $\sii^2$ the variance of 
 data $f^o_i$, the choice $\mu_i= \sii^{-2}$,
gives rise to the square distance  between 
$|f^o\ram$ and $| g\ram \in 
 \BM$ as given by:
\be
|| |f^o \ram - |g \ram ||^2 = \lam f^o- g| f^o- g \ram=
\sum_{i=1}^M (f^o_i - g_i)^2 \frac{1}{\sii^2}.
\ee
The above distance is known to be optimal, in a maximum 
likelihood sense, if the data errors are Gaussian distributed 
\cite{sta}.\\ 
The space of the physical system is considered to 
be the Euclidean $N$-dimensional 
real space $\RN$. The standard orthogonal basis 
in $\RN$ will be indicated by vectors 
$\kn \spa n=1,\ldots,N$, 
so that every vector $|r \ra \in \RN$ 
is represented as: 
\be
|r \ra = \sum_{n=1}^N \bn r \ra \kn =\sum_{n=1}^N r_n \kn.
\ee
 For any two vectors $| v \ra$ and $| r \ra$  in $\RN$ 
 the inner product is defined as:
\be
\la v |r \ra = \sum_{n=1}^N \la v |n \ra 
 \la n |r \ra = \sum_{n=1}^N v_n r_n. 
\ee
Using the adopted vector notation, 
equations (\ref{sy}) are recast: 
\be
\kfo= \AO \kp
\label{dam}
\ee 
with 
\be
|\pp \ra= \sum_{n=1}^{N} \kn \bn  \pp \ra = 
\sum_{n=1}^{N} {p}_n^{\frac{1}{2}}
\kn 
\label{p2o}
\ee
and operator $\AO:\RN \to \BM$ given by
\be
\AO = \sum_{n=1}^{N} \kfn \bn.
\ee
Vectors  $\kfn \in \BM $ are defined 
in such a way that $\lam i | f_n \ram =f_{i,n}$, i.e., 
\be
\kfn= \sum_{i=1}^{M}|i \ram  \mu_i  \lam i | f_n \ram 
    = \sum_{i=1}^{M}  \mu_i f_{i,n} |i \ram.
\ee
In the line of \cite{pre1}, in order to determine the
MaxEnt $|\pp\ra$ distribution we consider
as constraint of the optimisation precess a subset of
$k$ equations (\ref{sy}) labelled by indices 
$l_j\spa j=1, \ldots, k$. 
This leads to the following expression 
for the distribution:
\be
|\ppk \ra =(\frac{1}{N} - \frac{1}{N}\sumk \bg l_j \ra _{\mu}
\,\,_{\mu} \la l_j |\lambda^k\ra _{\mu})
\sumn \kn + \sumk \AT | l_j \ra_{\mu}\,\,{_{\mu}}\la l_j |\lambda^k\ra 
{_{\mu}}.
\label{kp}
\ee
with
\be
\kg= \sumn \kfn  \equiv \sumn  \AO |n \ra.
\label{vg}
\ee
The superscript $k$  in  $|\ppk \ra$ given above 
indicates that the 
distribution is built out of $k$ constraints.
The Lagrange multiplier vector $|\lambda^{(k)}\ra$ is
determined
by the requirement that $|\ppk \ra$ predicts a complete data
vector $\kfe = \AO |\ppk \ra \in \BM$ minimising the
distance to the observed vector $\kfo$. This is actually the 
prescription  given in \cite{pre1}. Nevertheless, the fact that 
 here the distance is  defined with 
respect to a measure, which  we propose to be dependent on the 
experimental data, implies that the 
 formalism of \cite{pre1}
 needs to be adapted to this 
 requirement.  In subsequent sections we discuss 
how this can be achieved in an straightforward manner 
by means of a recursive biorthogonalization 
technique for computing the Lagrange 
  multipliers which determine $|\ppk \ra$. 

\subsection{Determination of Lagrange multipliers} 
In order to estimate the 
Lagrange multipliers determining (\ref{kp}) we 
minimise the distance between the prediction through 
the physical model and 
observed data. 
As discussed in \cite{pre1,pre2} this entails to 
determine the Lagrange multipliers as
\be
\sumk |\alpha_{l_j}\ram \lam l_j |\lamb^{(k)}\ram = 
\FOK  |\lamb^{(k)}\ram = \hat{P}_{V_k} \kfot,
\label{ela}
\ee
where we have denoted: 
$\FOK= \sumk |\alpha_{l_j} \ram  \la l_j |$,
with
\be
|\alpha_{l_j} \ra  = \sumn \kfn \bfn l_j \ra -
\frac{1}{N}\kg \bg l_j \ra . \label{va}
\ee
Vector $\kfot$ is obtained from the data vector 
as $\kfot= \kfo- \frac{\kg}{N}$ and  $\hat{P}_{V_k}$ 
is the orthogonal projector onto the subspace spanned by 
$|\alpha_{l_j} \ram \;;\; j=1,\ldots,k$.  Here we wish  
this projector to account for the different weights of the 
data. This  will be achieved by recurse to a  
biorthogonalization technique \cite{fobio} which, as 
applied in this context,    
 produces biorthogonal vectors 
dependent on the weight assigned to each data.\\ 
Given a set of vectors
$|\alpha_{l_n}  \ram\;;\; n=1,\ldots,M$ we set 
 $|\psi_{l_1}\ram= |\alpha_{1}\ram$ and inductively 
 define vectors
$\kpsiktt$ as 
\be
\kpsiktt = \frac{\kpsik}{||\kpsik ||^2}
\ee
with 
\be
\kpsik =
|\alpha_{l_{k+1}}\ram - \op_{V_{k}} |\alpha_{l_{k+1}}\ram.
\ee
The dual vectors $ \balnkkts \spa n=1,\ldots,k+1$ 
which are obtained from  the recursive
equations 
\ben 
\balnkkts &=&\balnkts - \balnkts
             \alpha_{l_{k+1}}\ram {\bpsiktt}
             \sep n=1,\ldots,k\nonumber\\
\balkkkts &=& \frac{\bpsik}{\bpsik \alpha_{l_{k+1}} \ram }=
              \frac{\bpsik} {\bpsik  \psi_{k+1}\ram}= 
	     {\bpsiktt}, \label{rec}
\een 
satisfy the following properties

\begin{itemize}
\item a)  are biorthogonal with respect to vectors $\kaln \spa
n=1,\ldots,k+1$,
i.e., \be \balnkkts \alpha_{l_m}\ram  =
\delta_{l_m,l_n}\;\;\;;\;\;\;n=1,\ldots,k+1
\;\;;\;\; m=1,\ldots,k+1
\label{bi}
\ee
\item b) provide a representation of the 
orthogonal projection
 operator onto $V_{k+1}$ as given by: 
\be
\OPKK = \sumkk  \kaln  
\balnkkts= \OPKKT =\sumkk \kalnkkts  \baln. 
\label{teu} 
\ee
\end{itemize}
The proof of a) and b) parallels that of \cite{fobio,chap}, 
for the case of the standard Euclidean measure. \\
It follows 
from (\ref{teu}) and (\ref{ela}) that 
the Lagrange multipliers  yielding  $|\ppkk\ra$ are 
 obtained according to the 
recursive relation
\ben 
\lam l_n |\lamb^{(k+1)}
\ram&=& \la l_n |\lamb^{(k)} \ram - \balnkts \alpha_{l_{k+1}}\ram
\lam l_{k+1}|\lamb^{(k+1)} \ram \sep n=1,\ldots,k \nonumber\\
\lam l_{k+1}|\lamb^{(k+1)} \ram&=&\bpsiktt \til{f}^o\ram,
\label{c2}
\een
with
$\lam l_1 |\lamb^{(1)} \ram= 
\frac{\lam \alpha_{l_1}| \til{f}^o \ram}{|||\alpha_{l_1}\ram||^2}\,.$\\
 In writing down the above equations  
 we confidently assume that the  indices   
$l_n \spa n=1,\ldots,k+1$ are given to us. 
Of course, we must choose them somehow. 
How? 
The question does not possess a unique suitable answer, though.
 We tackle this problem below.

\section{Selection of indices}
The problem of deciding on the indices 
$l_n \spa n=1,\ldots,k$
to be considered in the construction of the
$|\ppk \ra$ distribution is far from be
a simple one. One would like, of course,
to choose the smallest set of  indices allowing
to minimise the distance between the observed
vector and the physical model. Unfortunately,
as already mentioned the search for a global minimum
is an NP-hard problem in most cases. A sensible
simplification is obtained by
resigning the goal of global minimisation and
accepting a less ambitious suboptimal solution which arises
from the following iterative procedure: At each iteration
the  indices  obtained in the previous steps are fixed, and
a new index is chosen so as to minimise the distance
between the data vector and the vector predicted by the
physical model. This is basically the strategy of
the forward selection approach  proposed in
\cite{pre1,pre2}. Such strategy, useful indeed
in many situations, is just
one among the many possible suboptimal strategies
that  one can envisage.
Here we advance a new approach which is
built out of two main ingredients:
i)A data independent technique for selecting constraints
to be discussed in section \ref{dre}, and
ii)A backward selection approach for reducing the
number of parameters of a given distribution.
To address the latter we need a technique
evolving in the reverse direction with respect
the forward technique of \cite{pre1,pre2}.
In this case the two
challenges we have to face are: a)The one
of deciding on the parameters
to be eliminated
b)The one of appropiertely modifying
the parameters one wishes to retain.
These two points are addressed in section \ref{redu}
by recurse to a backward birthogonalization
 approach. Before advancing the new strategy we
 would like to illustrate how
the forward selection approach of \cite{pre1,pre2},
can be adapted in an straightforward manner in order to
make it  suitable  when dealing with very noisy
data. This is the subject of the section \ref{sele}.

\subsection{Data dependent selection criterion}
\label{sele}
As proposed in \cite{pre1,pre2} a set of sub-indices 
$l_n \spa n=1,\ldots,k+1$ can be iteratively determined by 
selecting, at iteration $k+1$, the index $l_{k+1}$ corresponding 
to a vector $|\alpha_{l_{k+1}}\ram $ (Cf. Eq.(\ref{va}))
 that minimises the norm of the residual  resulting when 
approximating the observed data by the  physical  model.
This process is tantamount to selecting the index $l_{k+1}$
that  maximizes the functionals \cite{pre1}:
\be
e_n=|\lam \til{\psi}_n \kfot|^2 \sep  n=1,\ldots,M,
\label{crid}
\ee
with 
$|\til{\psi}_n\ram=\frac{|\psi_n\ram}{|||\psi_n\ram||}$
and $|{\psi}_n\ram =|\alpha_n\ram -\OPK |\alpha_n \ram$.\\
At this point, we would like to illustrate  the 
advantage of allowing different weights for each 
data.
We use the same example as in 
\cite{pre1} i.e.,  the data are generated as: 
\be
f_i^o= \sum_{n=1}^{50} p_n f_{i,n} + \epsilon_i
\;\;\;\;\; ;\;\;\;\;\; i=1,\ldots,100, \label{syn}
\ee
with $p_n$ represented by the continuous line of Figure 1, 
and $f_{i,n}=\exp(-n x_i) \spa x_i=0.01*i \spa i=1,\ldots,100
\spa n=1,\ldots,50.$  This is an extremely
bad conditioned problem. In  
 order to have a good approximation of the 
distribution of Figure 1, it was assumed in \cite{pre1} 
that we know the data within an uncertainty of $0.1\%$. 
Here we consider the errors to be much larger.
Each data is distorted by a zero mean Gaussian distributed 
random variable  of variance $\sii^{2}$ corresponding to 
$20\%$ of the data value. If, as in \cite{pre1}, we consider 
an uniform measure ($\mu=1$) the approximation we obtain 
 is represented by the dotted lines of Figure 1a 
 (for  2 different realizations of the data). 
 As we clearly gather from Figure 1b,
 by considering a nonuniform measure
 given as $\mu_i= \sii^{-2}\spa i=1,\ldots,100$ 
 the
 approximation is enormously improved 
 and becomes  stable  
against different realization of the data.\\ 
\subsection{Data independent selection criterion}
\label{dre}
This alternative criterion for selecting indices
is independent on the actual data. It
is meant to speed up the posterior selection
process and is grounded on the fact that
redundant equations arise as a consequence of
physical model. Hence, redundancy can be detected
without the actual realization of the
experimental measurements. In our
formalism each constraint, say the $l_k$-one is associated to
a vector $|\alpha_{l_k} \ram$. Hence the problem
of discriminating linearly independent constraints
is equivalent to the problem of discriminating linearly
independent vectors.  We address this problem by
recourse to a recently
introduce technique \cite{dre}, which allows for a
hierarchical selection giving rise to a stable
inverse problem. The goal is 
achieved by selecting, at each step, the index 
$l_k$ maximising the ratios: 
\be
r_n=\frac{|| |\psi_n \ram||^2}{ |||\alpha_n \ram||^2} \sep 
n=1\ldots,M. 
\label{recri}
\ee
This data independent technique for eliminating redundancy 
makes the posterior data processing 
much faster, as the selection of  indices for constructing 
the distribution can be carried out 
only on those  indices rendering independent vectors. There 
is also room for different post-processing strategies 
because, specially when the 
data are very noisy, the number of 
required Lagrange multipliers 
happens to be  smaller than the number of  indices 
rendering `numerical independence'. One possibility 
is to apply 
 the selection criterion discussed in the previous 
section,  but only on the preselected indices. 
Additional reduction of Lagrange Multipliers 
is made possible by a backward strategy to be introduced 
in the next section. 

\subsection{Reducing Lagrange Multipliers}
\label{redu}
As already discussed, the fact that Lagrange Multipliers 
are associated to constraints that are selected 
on a step by step basis  implies that at 
the end of the selection precess 
some Lagrange Multipliers  may have diminished relevance. 
To be in a position to eliminate Lagrange Multiplier 
of little relevance we need to develop an 
appropriate technique. 

Consider that we wish to reduce the number $k$ of 
Lagrange multipliers characterising a $|\ppk \ra$ 
distribution. 
Even if we know 
which particular parameters should be disregarded 
the actual process of removing them yields 
a non-linear  problem. The non-linearity follows from 
(\ref{ela}) where the Lagrange multipliers  in the 
left hand side of the equations are the coefficients 
of a linear  superposition of non-orthogonal vectors. 
The right hand side indicates that such a superposition  
is the orthogonal projection 
of  the vector $\kfot$ onto the subspace generated 
by vectors $|\alpha_{l_j} \ram  \spa j=1,\ldots,k$. 
Thus, within the framework of this Communication, 
the decision of eliminating some Lagrange multipliers  
comes along with the aim of
leaving the vector orthogonal projection  
onto the reduced subspace. This entails that 
we must recalculate the remaining Lagrange multipliers. 
The need for recalculating coefficients of a non-orthogonal 
linear expansion, when eliminating some others, is 
discussed in \cite{bab} where a backward biorthogonalization 
approach is advanced.  Such a technique, that we  
describe next, 
has been devised in order to modify biorthogonal 
vectors so as to  appropriately 
represent the orthogonal projector onto a reduced 
subspace.\\
Let us recall that $\vn= \spann \{|\alpha_{l_1} \ram, 
\ldots, \alpha_{l_k} \ram\}$ and  let 
$\vnj$ denote the subspace which is left by removing
the vector $\kalj$ from $\vn$, i.e,
\be
\vnj= \spann \{|\alpha_{l_1} \ram,  \ldots, |\alpha_{l_{j-1}} \ram, |\alpha_{l_{j+1}} \ram, \ldots, |\alpha_{l_k} \ram\}.
\ee
We have already discussed how to construct  
the orthogonal projector onto 
$\vn$ (Cf. Eq. (\ref{teu})). 
In order to 
represent the orthogonal projector onto the reduced 
subspace $\vnj$ the corresponding biorthogonal vectors  
 $\kalnkts$  need to be modified as established by 
the following theorem.\\
{\bf{Theorem 1:}} 
Given a set of vectors $\kaltnN \spa n=1,\ldots, k$
biorthogonal to vectors $\kaln \spa n=1,\ldots, k$ and
yielding a representation of $\pvn$ as given in (\ref{teu}),
a new set of biorthogonal vectors
$\kaltnNj \spa n=1,\ldots, j-1, j+1, \ldots,k$
yielding a representation of $\pvnj$ as given by
\be
\pvnj= \sumnj \kaln  \baltnNj=  \sumnj \kaltnNj  \baln.  
\label{teuj}
\ee
can be obtained from  vectors 
$\kaltnN \spa n=1,\ldots, k$ through 
the following equations:
\be
\kaltnNj = \kaltnN - \frac{\kaltjN \baltjN \altt_{l_n}^k \ram}{||\kaltjN||^2}
\sep  n=1,\ldots, j-1, j+1, \ldots,k.
\label{rec2}
\ee
The proof of this Theorem, as well as the proof of 
the Corollary 2 below,  
are  given in \cite{chap,bab}.\\
{\bf{Corollary 1:}} Let the Lagrange multiplier vector 
$|\lamb^k\ram$ satisfying (\ref{ela}) be given. 
Then, the Lagrange  multiplier vector  $|\lamb^{\Nj}\ram$ 
giving rise to the orthogonal projector onto the 
reduced subspace $\vnj$ is obtained from the previous 
$|\lamb^k\ram$ as follows:
\be
\lam l_n | \lamb^{\Nj} \ram= \lam l_n | \lamb^{k} \ram
 - \frac{\lam \altt_n^k \kaltjN \lam l_j |\lamb^{k}\ram}{||\kaltjN||^2}.
\label{core}
\ee
The proof trivially stems from (\ref{ela}) using 
(\ref{rec2})in (\ref{teuj}), since 
$\pvnj \kfot  =\sumnj \kaln   \lam l_n | \lamb^{\Nj}\ram $ implies
$ \baltnNj \til{f}^o \ram=  \lam l_n | \lamb^{\Nj} \ram 
\;\;\Box$\\
{\bf{Corollary 2:}} The following
relation between $|| \pvn \kfot ||$ and $||\pvnj \kfot||$ 
holds:
\be
|| \pvnj \kfot||^2 = ||   \pvn \kfot ||^2 - 
\frac{|\lam l_j| \lamb^k \ram|^2}{||\kaltjN||^2}.
\label{col}
\ee
Corollary 1 gives us a prescription to modify the 
Lagrange multipliers characterising a  $k$-parameters
distribution, 
if one of such multipliers is to be removed. Nevertheless, 
still the 
question has to be addressed as to how to choose 
the Lagrange multiplier to 
be disregarded. 
Corollary 2 suggests how the selection can be made optimal.
The following proposition is in order.\\
{\bf{Proposition 1:}} Let the Lagrange multipliers 
$ \lam l_n |\lamb^k\ram \spa n=1,\ldots,k$ 
and 
$\lam l_n |\lamb^{\Nj}\ram \spa n=1,\ldots, j-1,j+1,\ldots,k$ 
be obtained from
(\ref{ela}) and (\ref{core}) respectively. The 
Lagrange multiplier $ \lam l_j |\lamb^k\ram$ to be 
removed  for minimising the norm of the residual error
$|\Delta \ram= \pvn \kfot - \pvnj \kfot $ is the one 
yielding a minimum value of the quantities
\be
\frac{| \lam l_j | \lamb^k \ram|^2}{||\kaltjN||^2}
\sep j=1,\ldots M.
\label{crit}
\ee
{\bf{Proof}}: Since on the one hand 
$\pvn\pvnj=\pvnj \pvn=\pvnj$ and  on the oder hand 
orthogonal projectors are idempotent we have:
\be
|| \pvn \kfot - \pvnj \kfot ||^2 = 
\bfot \pvn \kfot - \bfot \pvnj \kfot = || \pvn \kfot ||^2 - || \pvnj \kfot||^2.
\ee
Making use of (\ref{col}), we further have
\be
|| \pvn \kfot - \pvnj \kfot||^2 = \frac{| \lam l_j | \lamb^k\ram|^2}{||\kaltjN||^2}.
\ee
It follows then that  $|| \pvn \kfot - \pvnj \kfot||^2$ is
minimum if $\frac{| \lam l_j | \lamb^k \ram|^2}
{||\kaltjN||^2}$ is minimum $\Box$\\
Successive applications of criterion (\ref{crit}) lead
to an algorithm for recursive backward approximations of the 
distribution. 
Indeed, let us assume that at the first iteration we eliminate
the $j$th-constraint yielding a minimum of (\ref{crit}).
We then construct the new reciprocal vectors 
(\ref{rec2}) and the corresponding new Lagrange multipliers
as prescribed in (\ref{core}). The process 
is to be stopped if the approximated distribution 
fails to predict the observed data within the required 
margin. 

{\subsection{Numerical example}}
We illustrate here an strategy consisting of the 
following steps: i)We use the data independent 
selection criterion for discriminating independent 
constraints.
ii)We apply the data dependent selection 
criterion on the previously selected indices. 
iii)The number of Lagrange multipliers obtained 
at step ii) is  
reduced and the remaining multipliers re-computed. \\
We consider the example  described  below.\\
The physical model yielding the 
matrix elements $f_{i,n}$ is given by the 
Lorentzian decays:
\be
f_{i,n}= \frac{1}{1+0.01*(i-100-n)^2}\sep
i=1,\ldots, 700 \sep n=1,\ldots,450.
\ee
We construct $700$ vectors 
$|\alpha_n \ram \spa n=1,\ldots,700$ 
as prescribed in (\ref{va}) and select 
 indices corresponding to the linearly 
independent vectors by the above descried 
technique for eliminating redundancy.  
Out of the redundant set of $700$ vectors we found 
100 linearly independent ones, up to a good 
 precision, which is assessed by the biorthogonality 
 quality of the corresponding basis and its reciprocal (dual).\\
The experimental measures were generated considering that
the distribution characterising the physical system 
is the sum of 5 Gaussian functions
represented by the continuous line of Figure 2. 
Each data was distorted  by 
a random error of variance $\sii^{2}$  
corresponding to $10\%$ of the data value. 
A realization of these data is shown in Figure 3.
The inversion problem in this example is much more stable 
than the one of the previous example so that 
the results do not vary much by weighting the data.
Hence in order to illustrate this strategy we use 
an uniform measure in all the involved procedures. 
Out of the pre-selected linearly independent
vectors, by using the data dependent strategy, 
we selected between 8 and 12 (depending on the 
particular realization of the data) to be able to 
predict the 700 pieces of data within the uncertainty
up to which the data were generate, i.e., we require that 
$||\kfe -\kfo||^ 2 <|| |\epsilon \ram ||^2$ where 
$|\epsilon \ram $ is a 
vector of components $\epsilon_i= t \sii$  
where, in general,  
$t$ is real number in the interval $[1 \,,\, 3]$. 
In this case we  first set $t=1.1$ 
The approximation 
of the corresponding distribution  
is depicted by the dotted lines 
of Figure 2a (for 5 different realization of the data). 
We then increased the value of $t$ up to 
$t=2$ and applied the proposed strategy for reducing 
Lagrange multipliers. In spite of the fact that 
the number of parameters was significantly reduced,  
(only 5 were kept) as it can be seen in Figure 2b 
the distribution is still a good approximation of 
the original one.
The inference  to the the data by this distribution 
is also of great quality. 
As shown in Figure 4 the predicted date are really close
to the  noiseless ones.  Notice that, by recourse to our approach,
we  are able to
de-noise and compress 700 data by using only 
$5$ Lagrange Multiplies.  
\section{Conclusions}
In this paper we have considered the problem of constructing
the $q=1/2$ MaxEnt distribution from redundant and noisy data.
A previously developed  approach has been generalised here
in order to
be able to incorporate, in a straightforward manner,
{\it information on the data errors}. The advantage
of this generalised approach, when dealing with
very noisy data, has been illustrated by a numerical
simulation.\\
Additionally, a new strategy for selecting relevant
constraints has been advanced.
The corresponding implementation consists of two different
steps. The first step is independent of the actual data, as
it operates by discriminating independent equations
{\it on the basis of the physical model}. The
data are used, a posteriori, to reduce further the number
of constraints. The latter process is carried out through a
 forward and backward procedure as follows:
First the selection is made starting from an initial
constraint and incorporating others, one by one, till
the observed data are predicted within a predetermined
precision.
Afterwards, the number
of parameters of the distribution is reduced
further by applying a backward selection criterion for
eliminating some of the Lagrange multipliers and recalculating
the remaining ones. It should be stressed that the
combination of the forward and backward procedures is not,
in general, equivalent to stopping the forward approach
at a corresponding earlier stage. The irreversibility
of the process is a consequence of the fact that, due to
the complexity of the problem,
the implementation of a selection criterion aiming at
global optimisation is not possible. The strategies we have
presented here are only optimal at each operational step.
Hence, they do not generate reversible procedures.\\
Considering the complexity of the mathematical problem
which is posed by the aim of constructing,
in an optimal way, the $q=1/2$ MaxEnt
distribution
from redundant and noisy constraints,
we believe that
the well founded suboptimal strategies we have employed here
should be of utility in a broad range of situations.

\section*{Acknowledgements}
Support from EPSRC (GR$/$R86355$/$01) is acknowledged.

\newpage
\begin{figure}[h]
\begin{center}
\input{exa1.tex}
\end{center}
\vspace{3cm}
{Figure. 1a: The theoretical distribution is represented by
the solid line. Each dotted line corresponds to 
the approximation we obtain by using an uniform measure 
($\mu=1$) for 2 different realization of the data.} 
\end{figure}

\newpage
\begin{figure}[h]
\begin{center}
\input{exa1p.tex}
\end{center}
\vspace{3cm}
{Figure 1b: The theoretical distribution is represented by
the solid line.  Each dotted line corresponds to 
the approximation
we obtain (for 5 different realization of the experiment)
by weighting  each data with a measure $\mu_i=\sii^{-2}$.}
\end{figure}

\newpage
\begin{figure}[h]
\begin{center}
\input{exa2.tex}
\end{center}
\vspace{3cm}
{Figure. 2a: The theoretical distribution is represented by
the solid line. The dotted lines correspond to 
the approximation we obtain  for 5 different realisations 
of the data.  Each line is constructed  by
iteratively selecting constraints out of the reduced set 
obtained by the data independent technique.}
\end{figure}

\newpage
\begin{figure}[h]
\begin{center}
\input{exa2b.tex}
\end{center}
\vspace{3cm}
{Figure. 2b: 
The theoretical distribution is represented by
the solid line. Each dotted line represents the approximation 
of the corresponding one in Figure 2a, after the elimination 
of some parameters.}
\end{figure}

\newpage
\begin{figure}[h]
\begin{center}
\input{exa2noi.tex}
\end{center}
\vspace{3cm}
{Figure 3: The simulated data after  distortion
by random noise.}
\end{figure}

\newpage
\begin{figure}[h]
\begin{center}
\input{exa2dat.tex}
\end{center}
\vspace{3cm}
{Figure 4: The theoretical data are represented by 
the continuous line. The dotted line corresponds of the 
predictions obtained by means of the approximation 
of Figure 2b.}
\end{figure}

\end{document}

%% file: exa1.tex
\setlength{\unitlength}{0.240900pt}
\ifx\plotpoint\undefined\newsavebox{\plotpoint}\fi
\sbox{\plotpoint}{\rule[-0.200pt]{0.400pt}{0.400pt}}%
\begin{picture}(1500,900)(0,0)
\font\gnuplot=cmr10 at 10pt
\gnuplot
\sbox{\plotpoint}{\rule[-0.200pt]{0.400pt}{0.400pt}}%
\put(201.0,123.0){\rule[-0.200pt]{4.818pt}{0.400pt}}
\put(181,123){\makebox(0,0)[r]{-0.04}}
\put(1419.0,123.0){\rule[-0.200pt]{4.818pt}{0.400pt}}
\put(201.0,205.0){\rule[-0.200pt]{4.818pt}{0.400pt}}
\put(181,205){\makebox(0,0)[r]{-0.03}}
\put(1419.0,205.0){\rule[-0.200pt]{4.818pt}{0.400pt}}
\put(201.0,287.0){\rule[-0.200pt]{4.818pt}{0.400pt}}
\put(181,287){\makebox(0,0)[r]{-0.02}}
\put(1419.0,287.0){\rule[-0.200pt]{4.818pt}{0.400pt}}
\put(201.0,369.0){\rule[-0.200pt]{4.818pt}{0.400pt}}
\put(181,369){\makebox(0,0)[r]{-0.01}}
\put(1419.0,369.0){\rule[-0.200pt]{4.818pt}{0.400pt}}
\put(201.0,451.0){\rule[-0.200pt]{4.818pt}{0.400pt}}
\put(181,451){\makebox(0,0)[r]{ 0}}
\put(1419.0,451.0){\rule[-0.200pt]{4.818pt}{0.400pt}}
\put(201.0,532.0){\rule[-0.200pt]{4.818pt}{0.400pt}}
\put(181,532){\makebox(0,0)[r]{ 0.01}}
\put(1419.0,532.0){\rule[-0.200pt]{4.818pt}{0.400pt}}
\put(201.0,614.0){\rule[-0.200pt]{4.818pt}{0.400pt}}
\put(181,614){\makebox(0,0)[r]{ 0.02}}
\put(1419.0,614.0){\rule[-0.200pt]{4.818pt}{0.400pt}}
\put(201.0,696.0){\rule[-0.200pt]{4.818pt}{0.400pt}}
\put(181,696){\makebox(0,0)[r]{ 0.03}}
\put(1419.0,696.0){\rule[-0.200pt]{4.818pt}{0.400pt}}
\put(201.0,778.0){\rule[-0.200pt]{4.818pt}{0.400pt}}
\put(181,778){\makebox(0,0)[r]{ 0.04}}
\put(1419.0,778.0){\rule[-0.200pt]{4.818pt}{0.400pt}}
\put(201.0,860.0){\rule[-0.200pt]{4.818pt}{0.400pt}}
\put(181,860){\makebox(0,0)[r]{ 0.05}}
\put(1419.0,860.0){\rule[-0.200pt]{4.818pt}{0.400pt}}
\put(302.0,123.0){\rule[-0.200pt]{0.400pt}{4.818pt}}
\put(302,82){\makebox(0,0){ 5}}
\put(302.0,840.0){\rule[-0.200pt]{0.400pt}{4.818pt}}
\put(428.0,123.0){\rule[-0.200pt]{0.400pt}{4.818pt}}
\put(428,82){\makebox(0,0){ 10}}
\put(428.0,840.0){\rule[-0.200pt]{0.400pt}{4.818pt}}
\put(555.0,123.0){\rule[-0.200pt]{0.400pt}{4.818pt}}
\put(555,82){\makebox(0,0){ 15}}
\put(555.0,840.0){\rule[-0.200pt]{0.400pt}{4.818pt}}
\put(681.0,123.0){\rule[-0.200pt]{0.400pt}{4.818pt}}
\put(681,82){\makebox(0,0){ 20}}
\put(681.0,840.0){\rule[-0.200pt]{0.400pt}{4.818pt}}
\put(807.0,123.0){\rule[-0.200pt]{0.400pt}{4.818pt}}
\put(807,82){\makebox(0,0){ 25}}
\put(807.0,840.0){\rule[-0.200pt]{0.400pt}{4.818pt}}
\put(934.0,123.0){\rule[-0.200pt]{0.400pt}{4.818pt}}
\put(934,82){\makebox(0,0){ 30}}
\put(934.0,840.0){\rule[-0.200pt]{0.400pt}{4.818pt}}
\put(1060.0,123.0){\rule[-0.200pt]{0.400pt}{4.818pt}}
\put(1060,82){\makebox(0,0){ 35}}
\put(1060.0,840.0){\rule[-0.200pt]{0.400pt}{4.818pt}}
\put(1186.0,123.0){\rule[-0.200pt]{0.400pt}{4.818pt}}
\put(1186,82){\makebox(0,0){ 40}}
\put(1186.0,840.0){\rule[-0.200pt]{0.400pt}{4.818pt}}
\put(1313.0,123.0){\rule[-0.200pt]{0.400pt}{4.818pt}}
\put(1313,82){\makebox(0,0){ 45}}
\put(1313.0,840.0){\rule[-0.200pt]{0.400pt}{4.818pt}}
\put(1439.0,123.0){\rule[-0.200pt]{0.400pt}{4.818pt}}
\put(1439,82){\makebox(0,0){ 50}}
\put(1439.0,840.0){\rule[-0.200pt]{0.400pt}{4.818pt}}
\put(201.0,123.0){\rule[-0.200pt]{298.234pt}{0.400pt}}
\put(1439.0,123.0){\rule[-0.200pt]{0.400pt}{177.543pt}}
\put(201.0,860.0){\rule[-0.200pt]{298.234pt}{0.400pt}}
\put(40,791){\makebox(0,0){$p_n$}}
\put(1360,21){\makebox(0,0){$n$}}
\put(201.0,123.0){\rule[-0.200pt]{0.400pt}{177.543pt}}
\put(201,524){\usebox{\plotpoint}}
\multiput(201.58,524.00)(0.497,1.797){47}{\rule{0.120pt}{1.524pt}}
\multiput(200.17,524.00)(25.000,85.837){2}{\rule{0.400pt}{0.762pt}}
\multiput(226.58,613.00)(0.497,1.142){49}{\rule{0.120pt}{1.008pt}}
\multiput(225.17,613.00)(26.000,56.908){2}{\rule{0.400pt}{0.504pt}}
\multiput(252.58,672.00)(0.497,0.681){47}{\rule{0.120pt}{0.644pt}}
\multiput(251.17,672.00)(25.000,32.663){2}{\rule{0.400pt}{0.322pt}}
\multiput(277.00,706.58)(0.625,0.496){37}{\rule{0.600pt}{0.119pt}}
\multiput(277.00,705.17)(23.755,20.000){2}{\rule{0.300pt}{0.400pt}}
\multiput(302.00,726.59)(1.427,0.489){15}{\rule{1.211pt}{0.118pt}}
\multiput(302.00,725.17)(22.486,9.000){2}{\rule{0.606pt}{0.400pt}}
\put(327,735.17){\rule{5.300pt}{0.400pt}}
\multiput(327.00,734.17)(15.000,2.000){2}{\rule{2.650pt}{0.400pt}}
\put(353,735.17){\rule{5.100pt}{0.400pt}}
\multiput(353.00,736.17)(14.415,-2.000){2}{\rule{2.550pt}{0.400pt}}
\multiput(378.00,733.94)(3.552,-0.468){5}{\rule{2.600pt}{0.113pt}}
\multiput(378.00,734.17)(19.604,-4.000){2}{\rule{1.300pt}{0.400pt}}
\multiput(403.00,729.93)(1.865,-0.485){11}{\rule{1.529pt}{0.117pt}}
\multiput(403.00,730.17)(21.827,-7.000){2}{\rule{0.764pt}{0.400pt}}
\multiput(428.00,722.93)(1.942,-0.485){11}{\rule{1.586pt}{0.117pt}}
\multiput(428.00,723.17)(22.709,-7.000){2}{\rule{0.793pt}{0.400pt}}
\multiput(454.00,715.93)(1.616,-0.488){13}{\rule{1.350pt}{0.117pt}}
\multiput(454.00,716.17)(22.198,-8.000){2}{\rule{0.675pt}{0.400pt}}
\multiput(479.00,707.93)(1.427,-0.489){15}{\rule{1.211pt}{0.118pt}}
\multiput(479.00,708.17)(22.486,-9.000){2}{\rule{0.606pt}{0.400pt}}
\multiput(504.00,698.93)(1.427,-0.489){15}{\rule{1.211pt}{0.118pt}}
\multiput(504.00,699.17)(22.486,-9.000){2}{\rule{0.606pt}{0.400pt}}
\multiput(529.00,689.93)(1.682,-0.488){13}{\rule{1.400pt}{0.117pt}}
\multiput(529.00,690.17)(23.094,-8.000){2}{\rule{0.700pt}{0.400pt}}
\multiput(555.00,681.93)(1.427,-0.489){15}{\rule{1.211pt}{0.118pt}}
\multiput(555.00,682.17)(22.486,-9.000){2}{\rule{0.606pt}{0.400pt}}
\multiput(580.00,672.93)(1.616,-0.488){13}{\rule{1.350pt}{0.117pt}}
\multiput(580.00,673.17)(22.198,-8.000){2}{\rule{0.675pt}{0.400pt}}
\multiput(605.00,664.93)(1.682,-0.488){13}{\rule{1.400pt}{0.117pt}}
\multiput(605.00,665.17)(23.094,-8.000){2}{\rule{0.700pt}{0.400pt}}
\multiput(631.00,656.93)(1.616,-0.488){13}{\rule{1.350pt}{0.117pt}}
\multiput(631.00,657.17)(22.198,-8.000){2}{\rule{0.675pt}{0.400pt}}
\multiput(656.00,648.93)(1.865,-0.485){11}{\rule{1.529pt}{0.117pt}}
\multiput(656.00,649.17)(21.827,-7.000){2}{\rule{0.764pt}{0.400pt}}
\multiput(681.00,641.93)(1.865,-0.485){11}{\rule{1.529pt}{0.117pt}}
\multiput(681.00,642.17)(21.827,-7.000){2}{\rule{0.764pt}{0.400pt}}
\multiput(706.00,634.93)(1.942,-0.485){11}{\rule{1.586pt}{0.117pt}}
\multiput(706.00,635.17)(22.709,-7.000){2}{\rule{0.793pt}{0.400pt}}
\multiput(732.00,627.93)(2.208,-0.482){9}{\rule{1.767pt}{0.116pt}}
\multiput(732.00,628.17)(21.333,-6.000){2}{\rule{0.883pt}{0.400pt}}
\multiput(757.00,621.93)(1.865,-0.485){11}{\rule{1.529pt}{0.117pt}}
\multiput(757.00,622.17)(21.827,-7.000){2}{\rule{0.764pt}{0.400pt}}
\multiput(782.00,614.93)(2.208,-0.482){9}{\rule{1.767pt}{0.116pt}}
\multiput(782.00,615.17)(21.333,-6.000){2}{\rule{0.883pt}{0.400pt}}
\multiput(807.00,608.93)(2.825,-0.477){7}{\rule{2.180pt}{0.115pt}}
\multiput(807.00,609.17)(21.475,-5.000){2}{\rule{1.090pt}{0.400pt}}
\multiput(833.00,603.93)(2.208,-0.482){9}{\rule{1.767pt}{0.116pt}}
\multiput(833.00,604.17)(21.333,-6.000){2}{\rule{0.883pt}{0.400pt}}
\multiput(858.00,597.93)(2.714,-0.477){7}{\rule{2.100pt}{0.115pt}}
\multiput(858.00,598.17)(20.641,-5.000){2}{\rule{1.050pt}{0.400pt}}
\multiput(883.00,592.93)(2.714,-0.477){7}{\rule{2.100pt}{0.115pt}}
\multiput(883.00,593.17)(20.641,-5.000){2}{\rule{1.050pt}{0.400pt}}
\multiput(908.00,587.93)(2.825,-0.477){7}{\rule{2.180pt}{0.115pt}}
\multiput(908.00,588.17)(21.475,-5.000){2}{\rule{1.090pt}{0.400pt}}
\multiput(934.00,582.93)(2.714,-0.477){7}{\rule{2.100pt}{0.115pt}}
\multiput(934.00,583.17)(20.641,-5.000){2}{\rule{1.050pt}{0.400pt}}
\multiput(959.00,577.94)(3.552,-0.468){5}{\rule{2.600pt}{0.113pt}}
\multiput(959.00,578.17)(19.604,-4.000){2}{\rule{1.300pt}{0.400pt}}
\multiput(984.00,573.94)(3.552,-0.468){5}{\rule{2.600pt}{0.113pt}}
\multiput(984.00,574.17)(19.604,-4.000){2}{\rule{1.300pt}{0.400pt}}
\multiput(1009.00,569.94)(3.698,-0.468){5}{\rule{2.700pt}{0.113pt}}
\multiput(1009.00,570.17)(20.396,-4.000){2}{\rule{1.350pt}{0.400pt}}
\multiput(1035.00,565.94)(3.552,-0.468){5}{\rule{2.600pt}{0.113pt}}
\multiput(1035.00,566.17)(19.604,-4.000){2}{\rule{1.300pt}{0.400pt}}
\multiput(1060.00,561.94)(3.552,-0.468){5}{\rule{2.600pt}{0.113pt}}
\multiput(1060.00,562.17)(19.604,-4.000){2}{\rule{1.300pt}{0.400pt}}
\multiput(1085.00,557.95)(5.597,-0.447){3}{\rule{3.567pt}{0.108pt}}
\multiput(1085.00,558.17)(18.597,-3.000){2}{\rule{1.783pt}{0.400pt}}
\multiput(1111.00,554.95)(5.374,-0.447){3}{\rule{3.433pt}{0.108pt}}
\multiput(1111.00,555.17)(17.874,-3.000){2}{\rule{1.717pt}{0.400pt}}
\multiput(1136.00,551.94)(3.552,-0.468){5}{\rule{2.600pt}{0.113pt}}
\multiput(1136.00,552.17)(19.604,-4.000){2}{\rule{1.300pt}{0.400pt}}
\multiput(1161.00,547.95)(5.374,-0.447){3}{\rule{3.433pt}{0.108pt}}
\multiput(1161.00,548.17)(17.874,-3.000){2}{\rule{1.717pt}{0.400pt}}
\multiput(1186.00,544.95)(5.597,-0.447){3}{\rule{3.567pt}{0.108pt}}
\multiput(1186.00,545.17)(18.597,-3.000){2}{\rule{1.783pt}{0.400pt}}
\put(1212,541.17){\rule{5.100pt}{0.400pt}}
\multiput(1212.00,542.17)(14.415,-2.000){2}{\rule{2.550pt}{0.400pt}}
\multiput(1237.00,539.95)(5.374,-0.447){3}{\rule{3.433pt}{0.108pt}}
\multiput(1237.00,540.17)(17.874,-3.000){2}{\rule{1.717pt}{0.400pt}}
\multiput(1262.00,536.95)(5.374,-0.447){3}{\rule{3.433pt}{0.108pt}}
\multiput(1262.00,537.17)(17.874,-3.000){2}{\rule{1.717pt}{0.400pt}}
\put(1287,533.17){\rule{5.300pt}{0.400pt}}
\multiput(1287.00,534.17)(15.000,-2.000){2}{\rule{2.650pt}{0.400pt}}
\multiput(1313.00,531.95)(5.374,-0.447){3}{\rule{3.433pt}{0.108pt}}
\multiput(1313.00,532.17)(17.874,-3.000){2}{\rule{1.717pt}{0.400pt}}
\put(1338,528.17){\rule{5.100pt}{0.400pt}}
\multiput(1338.00,529.17)(14.415,-2.000){2}{\rule{2.550pt}{0.400pt}}
\put(1363,526.17){\rule{5.100pt}{0.400pt}}
\multiput(1363.00,527.17)(14.415,-2.000){2}{\rule{2.550pt}{0.400pt}}
\put(1388,524.17){\rule{5.300pt}{0.400pt}}
\multiput(1388.00,525.17)(15.000,-2.000){2}{\rule{2.650pt}{0.400pt}}
\put(1414,522.17){\rule{5.100pt}{0.400pt}}
\multiput(1414.00,523.17)(14.415,-2.000){2}{\rule{2.550pt}{0.400pt}}
\sbox{\plotpoint}{\rule[-0.500pt]{1.000pt}{1.000pt}}%
\put(201,552){\usebox{\plotpoint}}
\multiput(201,552)(12.778,16.356){2}{\usebox{\plotpoint}}
\multiput(226,584)(13.593,15.685){2}{\usebox{\plotpoint}}
\multiput(252,614)(14.101,15.230){2}{\usebox{\plotpoint}}
\multiput(277,641)(14.676,14.676){2}{\usebox{\plotpoint}}
\put(312.20,674.97){\usebox{\plotpoint}}
\multiput(327,688)(16.147,13.041){2}{\usebox{\plotpoint}}
\multiput(353,709)(16.844,12.128){2}{\usebox{\plotpoint}}
\put(394.71,737.69){\usebox{\plotpoint}}
\put(412.52,748.33){\usebox{\plotpoint}}
\multiput(428,757)(19.115,8.087){2}{\usebox{\plotpoint}}
\put(469.12,774.05){\usebox{\plotpoint}}
\put(488.52,781.43){\usebox{\plotpoint}}
\multiput(504,787)(20.182,4.844){2}{\usebox{\plotpoint}}
\put(548.87,796.06){\usebox{\plotpoint}}
\put(569.45,798.73){\usebox{\plotpoint}}
\put(590.12,800.40){\usebox{\plotpoint}}
\put(610.86,800.77){\usebox{\plotpoint}}
\multiput(631,800)(20.689,-1.655){2}{\usebox{\plotpoint}}
\put(672.82,795.31){\usebox{\plotpoint}}
\put(693.23,791.55){\usebox{\plotpoint}}
\put(713.47,786.99){\usebox{\plotpoint}}
\multiput(732,782)(19.768,-6.326){2}{\usebox{\plotpoint}}
\put(772.83,768.30){\usebox{\plotpoint}}
\put(792.08,760.57){\usebox{\plotpoint}}
\multiput(807,754)(18.845,-8.698){2}{\usebox{\plotpoint}}
\put(848.37,734.01){\usebox{\plotpoint}}
\put(866.49,723.90){\usebox{\plotpoint}}
\multiput(883,714)(17.798,-10.679){2}{\usebox{\plotpoint}}
\put(919.60,691.41){\usebox{\plotpoint}}
\multiput(934,682)(16.844,-12.128){2}{\usebox{\plotpoint}}
\put(970.35,655.37){\usebox{\plotpoint}}
\multiput(984,645)(16.207,-12.966){2}{\usebox{\plotpoint}}
\put(1019.39,617.01){\usebox{\plotpoint}}
\multiput(1035,605)(15.581,-13.712){2}{\usebox{\plotpoint}}
\multiput(1060,583)(15.275,-14.053){2}{\usebox{\plotpoint}}
\put(1097.35,548.60){\usebox{\plotpoint}}
\multiput(1111,536)(14.973,-14.374){2}{\usebox{\plotpoint}}
\multiput(1136,512)(14.676,-14.676){2}{\usebox{\plotpoint}}
\multiput(1161,487)(14.101,-15.230){2}{\usebox{\plotpoint}}
\put(1200.08,445.92){\usebox{\plotpoint}}
\multiput(1212,434)(13.823,-15.482){2}{\usebox{\plotpoint}}
\multiput(1237,406)(13.823,-15.482){2}{\usebox{\plotpoint}}
\multiput(1262,378)(13.552,-15.720){2}{\usebox{\plotpoint}}
\multiput(1287,349)(13.593,-15.685){2}{\usebox{\plotpoint}}
\multiput(1313,319)(13.287,-15.945){2}{\usebox{\plotpoint}}
\put(1350.14,273.95){\usebox{\plotpoint}}
\multiput(1363,258)(12.778,-16.356){2}{\usebox{\plotpoint}}
\multiput(1388,226)(13.088,-16.109){2}{\usebox{\plotpoint}}
\multiput(1414,194)(12.778,-16.356){2}{\usebox{\plotpoint}}
\put(1439,162){\usebox{\plotpoint}}
\put(201,692){\usebox{\plotpoint}}
\multiput(201,692)(20.495,-3.279){2}{\usebox{\plotpoint}}
\put(242.00,685.54){\usebox{\plotpoint}}
\put(262.57,682.73){\usebox{\plotpoint}}
\put(283.14,680.02){\usebox{\plotpoint}}
\multiput(302,677)(20.608,-2.473){2}{\usebox{\plotpoint}}
\put(344.78,671.26){\usebox{\plotpoint}}
\put(365.35,668.52){\usebox{\plotpoint}}
\put(385.91,665.73){\usebox{\plotpoint}}
\multiput(403,663)(20.608,-2.473){2}{\usebox{\plotpoint}}
\put(447.65,657.73){\usebox{\plotpoint}}
\put(468.19,654.73){\usebox{\plotpoint}}
\put(488.73,651.83){\usebox{\plotpoint}}
\put(509.34,649.36){\usebox{\plotpoint}}
\multiput(529,647)(20.514,-3.156){2}{\usebox{\plotpoint}}
\put(571.05,641.07){\usebox{\plotpoint}}
\put(591.65,638.60){\usebox{\plotpoint}}
\put(612.26,636.16){\usebox{\plotpoint}}
\multiput(631,634)(20.608,-2.473){2}{\usebox{\plotpoint}}
\put(674.10,628.83){\usebox{\plotpoint}}
\put(694.71,626.36){\usebox{\plotpoint}}
\put(715.32,623.92){\usebox{\plotpoint}}
\multiput(732,622)(20.608,-2.473){2}{\usebox{\plotpoint}}
\put(777.15,616.58){\usebox{\plotpoint}}
\put(797.76,614.11){\usebox{\plotpoint}}
\put(818.37,611.69){\usebox{\plotpoint}}
\put(838.99,609.28){\usebox{\plotpoint}}
\multiput(858,607)(20.608,-2.473){2}{\usebox{\plotpoint}}
\put(900.81,601.86){\usebox{\plotpoint}}
\put(921.47,599.96){\usebox{\plotpoint}}
\put(942.13,598.02){\usebox{\plotpoint}}
\multiput(959,596)(20.608,-2.473){2}{\usebox{\plotpoint}}
\put(1003.96,590.61){\usebox{\plotpoint}}
\put(1024.63,588.80){\usebox{\plotpoint}}
\put(1045.28,586.77){\usebox{\plotpoint}}
\put(1065.89,584.29){\usebox{\plotpoint}}
\multiput(1085,582)(20.694,-1.592){2}{\usebox{\plotpoint}}
\put(1127.82,577.98){\usebox{\plotpoint}}
\put(1148.48,576.00){\usebox{\plotpoint}}
\put(1169.14,574.02){\usebox{\plotpoint}}
\multiput(1186,572)(20.694,-1.592){2}{\usebox{\plotpoint}}
\put(1231.07,567.71){\usebox{\plotpoint}}
\put(1251.73,565.82){\usebox{\plotpoint}}
\put(1272.38,563.75){\usebox{\plotpoint}}
\put(1293.01,561.54){\usebox{\plotpoint}}
\multiput(1313,560)(20.689,-1.655){2}{\usebox{\plotpoint}}
\put(1355.02,555.96){\usebox{\plotpoint}}
\put(1375.68,553.99){\usebox{\plotpoint}}
\put(1396.37,552.36){\usebox{\plotpoint}}
\multiput(1414,551)(20.608,-2.473){2}{\usebox{\plotpoint}}
\put(1439,548){\usebox{\plotpoint}}
\end{picture}

%% file: exa1p.tex
\setlength{\unitlength}{0.240900pt}
\ifx\plotpoint\undefined\newsavebox{\plotpoint}\fi
\sbox{\plotpoint}{\rule[-0.200pt]{0.400pt}{0.400pt}}%
\begin{picture}(1500,900)(0,0)
\font\gnuplot=cmr10 at 10pt
\gnuplot
\sbox{\plotpoint}{\rule[-0.200pt]{0.400pt}{0.400pt}}%
\put(221.0,123.0){\rule[-0.200pt]{4.818pt}{0.400pt}}
\put(201,123){\makebox(0,0)[r]{ 0.005}}
\put(1419.0,123.0){\rule[-0.200pt]{4.818pt}{0.400pt}}
\put(221.0,246.0){\rule[-0.200pt]{4.818pt}{0.400pt}}
\put(201,246){\makebox(0,0)[r]{ 0.01}}
\put(1419.0,246.0){\rule[-0.200pt]{4.818pt}{0.400pt}}
\put(221.0,369.0){\rule[-0.200pt]{4.818pt}{0.400pt}}
\put(201,369){\makebox(0,0)[r]{ 0.015}}
\put(1419.0,369.0){\rule[-0.200pt]{4.818pt}{0.400pt}}
\put(221.0,491.0){\rule[-0.200pt]{4.818pt}{0.400pt}}
\put(201,491){\makebox(0,0)[r]{ 0.02}}
\put(1419.0,491.0){\rule[-0.200pt]{4.818pt}{0.400pt}}
\put(221.0,614.0){\rule[-0.200pt]{4.818pt}{0.400pt}}
\put(201,614){\makebox(0,0)[r]{ 0.025}}
\put(1419.0,614.0){\rule[-0.200pt]{4.818pt}{0.400pt}}
\put(221.0,737.0){\rule[-0.200pt]{4.818pt}{0.400pt}}
\put(201,737){\makebox(0,0)[r]{ 0.03}}
\put(1419.0,737.0){\rule[-0.200pt]{4.818pt}{0.400pt}}
\put(221.0,860.0){\rule[-0.200pt]{4.818pt}{0.400pt}}
\put(201,860){\makebox(0,0)[r]{ 0.035}}
\put(1419.0,860.0){\rule[-0.200pt]{4.818pt}{0.400pt}}
\put(320.0,123.0){\rule[-0.200pt]{0.400pt}{4.818pt}}
\put(320,82){\makebox(0,0){ 5}}
\put(320.0,840.0){\rule[-0.200pt]{0.400pt}{4.818pt}}
\put(445.0,123.0){\rule[-0.200pt]{0.400pt}{4.818pt}}
\put(445,82){\makebox(0,0){ 10}}
\put(445.0,840.0){\rule[-0.200pt]{0.400pt}{4.818pt}}
\put(569.0,123.0){\rule[-0.200pt]{0.400pt}{4.818pt}}
\put(569,82){\makebox(0,0){ 15}}
\put(569.0,840.0){\rule[-0.200pt]{0.400pt}{4.818pt}}
\put(693.0,123.0){\rule[-0.200pt]{0.400pt}{4.818pt}}
\put(693,82){\makebox(0,0){ 20}}
\put(693.0,840.0){\rule[-0.200pt]{0.400pt}{4.818pt}}
\put(818.0,123.0){\rule[-0.200pt]{0.400pt}{4.818pt}}
\put(818,82){\makebox(0,0){ 25}}
\put(818.0,840.0){\rule[-0.200pt]{0.400pt}{4.818pt}}
\put(942.0,123.0){\rule[-0.200pt]{0.400pt}{4.818pt}}
\put(942,82){\makebox(0,0){ 30}}
\put(942.0,840.0){\rule[-0.200pt]{0.400pt}{4.818pt}}
\put(1066.0,123.0){\rule[-0.200pt]{0.400pt}{4.818pt}}
\put(1066,82){\makebox(0,0){ 35}}
\put(1066.0,840.0){\rule[-0.200pt]{0.400pt}{4.818pt}}
\put(1190.0,123.0){\rule[-0.200pt]{0.400pt}{4.818pt}}
\put(1190,82){\makebox(0,0){ 40}}
\put(1190.0,840.0){\rule[-0.200pt]{0.400pt}{4.818pt}}
\put(1315.0,123.0){\rule[-0.200pt]{0.400pt}{4.818pt}}
\put(1315,82){\makebox(0,0){ 45}}
\put(1315.0,840.0){\rule[-0.200pt]{0.400pt}{4.818pt}}
\put(1439.0,123.0){\rule[-0.200pt]{0.400pt}{4.818pt}}
\put(1439,82){\makebox(0,0){ 50}}
\put(1439.0,840.0){\rule[-0.200pt]{0.400pt}{4.818pt}}
\put(221.0,123.0){\rule[-0.200pt]{293.416pt}{0.400pt}}
\put(1439.0,123.0){\rule[-0.200pt]{0.400pt}{177.543pt}}
\put(221.0,860.0){\rule[-0.200pt]{293.416pt}{0.400pt}}
\put(40,791){\makebox(0,0){$p_n$}}
\put(1370,21){\makebox(0,0){$n$}}
\put(221.0,123.0){\rule[-0.200pt]{0.400pt}{177.543pt}}
\put(221,219){\usebox{\plotpoint}}
\multiput(221.58,219.00)(0.497,5.450){47}{\rule{0.120pt}{4.404pt}}
\multiput(220.17,219.00)(25.000,259.859){2}{\rule{0.400pt}{2.202pt}}
\multiput(246.58,488.00)(0.497,3.563){47}{\rule{0.120pt}{2.916pt}}
\multiput(245.17,488.00)(25.000,169.948){2}{\rule{0.400pt}{1.458pt}}
\multiput(271.58,664.00)(0.497,2.102){47}{\rule{0.120pt}{1.764pt}}
\multiput(270.17,664.00)(25.000,100.339){2}{\rule{0.400pt}{0.882pt}}
\multiput(296.58,768.00)(0.496,1.196){45}{\rule{0.120pt}{1.050pt}}
\multiput(295.17,768.00)(24.000,54.821){2}{\rule{0.400pt}{0.525pt}}
\multiput(320.58,825.00)(0.497,0.539){47}{\rule{0.120pt}{0.532pt}}
\multiput(319.17,825.00)(25.000,25.896){2}{\rule{0.400pt}{0.266pt}}
\multiput(345.00,852.59)(1.616,0.488){13}{\rule{1.350pt}{0.117pt}}
\multiput(345.00,851.17)(22.198,8.000){2}{\rule{0.675pt}{0.400pt}}
\multiput(370.00,858.93)(2.208,-0.482){9}{\rule{1.767pt}{0.116pt}}
\multiput(370.00,859.17)(21.333,-6.000){2}{\rule{0.883pt}{0.400pt}}
\multiput(395.00,852.92)(0.901,-0.494){25}{\rule{0.814pt}{0.119pt}}
\multiput(395.00,853.17)(23.310,-14.000){2}{\rule{0.407pt}{0.400pt}}
\multiput(420.00,838.92)(0.659,-0.495){35}{\rule{0.626pt}{0.119pt}}
\multiput(420.00,839.17)(23.700,-19.000){2}{\rule{0.313pt}{0.400pt}}
\multiput(445.00,819.92)(0.567,-0.496){41}{\rule{0.555pt}{0.120pt}}
\multiput(445.00,820.17)(23.849,-22.000){2}{\rule{0.277pt}{0.400pt}}
\multiput(470.58,796.86)(0.496,-0.519){45}{\rule{0.120pt}{0.517pt}}
\multiput(469.17,797.93)(24.000,-23.928){2}{\rule{0.400pt}{0.258pt}}
\multiput(494.00,772.92)(0.498,-0.497){47}{\rule{0.500pt}{0.120pt}}
\multiput(494.00,773.17)(23.962,-25.000){2}{\rule{0.250pt}{0.400pt}}
\multiput(519.58,746.86)(0.497,-0.519){47}{\rule{0.120pt}{0.516pt}}
\multiput(518.17,747.93)(25.000,-24.929){2}{\rule{0.400pt}{0.258pt}}
\multiput(544.58,720.86)(0.497,-0.519){47}{\rule{0.120pt}{0.516pt}}
\multiput(543.17,721.93)(25.000,-24.929){2}{\rule{0.400pt}{0.258pt}}
\multiput(569.00,695.92)(0.498,-0.497){47}{\rule{0.500pt}{0.120pt}}
\multiput(569.00,696.17)(23.962,-25.000){2}{\rule{0.250pt}{0.400pt}}
\multiput(594.00,670.92)(0.498,-0.497){47}{\rule{0.500pt}{0.120pt}}
\multiput(594.00,671.17)(23.962,-25.000){2}{\rule{0.250pt}{0.400pt}}
\multiput(619.00,645.92)(0.519,-0.496){45}{\rule{0.517pt}{0.120pt}}
\multiput(619.00,646.17)(23.928,-24.000){2}{\rule{0.258pt}{0.400pt}}
\multiput(644.00,621.92)(0.520,-0.496){43}{\rule{0.517pt}{0.120pt}}
\multiput(644.00,622.17)(22.926,-23.000){2}{\rule{0.259pt}{0.400pt}}
\multiput(668.00,598.92)(0.567,-0.496){41}{\rule{0.555pt}{0.120pt}}
\multiput(668.00,599.17)(23.849,-22.000){2}{\rule{0.277pt}{0.400pt}}
\multiput(693.00,576.92)(0.567,-0.496){41}{\rule{0.555pt}{0.120pt}}
\multiput(693.00,577.17)(23.849,-22.000){2}{\rule{0.277pt}{0.400pt}}
\multiput(718.00,554.92)(0.625,-0.496){37}{\rule{0.600pt}{0.119pt}}
\multiput(718.00,555.17)(23.755,-20.000){2}{\rule{0.300pt}{0.400pt}}
\multiput(743.00,534.92)(0.625,-0.496){37}{\rule{0.600pt}{0.119pt}}
\multiput(743.00,535.17)(23.755,-20.000){2}{\rule{0.300pt}{0.400pt}}
\multiput(768.00,514.92)(0.659,-0.495){35}{\rule{0.626pt}{0.119pt}}
\multiput(768.00,515.17)(23.700,-19.000){2}{\rule{0.313pt}{0.400pt}}
\multiput(793.00,495.92)(0.696,-0.495){33}{\rule{0.656pt}{0.119pt}}
\multiput(793.00,496.17)(23.639,-18.000){2}{\rule{0.328pt}{0.400pt}}
\multiput(818.00,477.92)(0.708,-0.495){31}{\rule{0.665pt}{0.119pt}}
\multiput(818.00,478.17)(22.620,-17.000){2}{\rule{0.332pt}{0.400pt}}
\multiput(842.00,460.92)(0.785,-0.494){29}{\rule{0.725pt}{0.119pt}}
\multiput(842.00,461.17)(23.495,-16.000){2}{\rule{0.363pt}{0.400pt}}
\multiput(867.00,444.92)(0.785,-0.494){29}{\rule{0.725pt}{0.119pt}}
\multiput(867.00,445.17)(23.495,-16.000){2}{\rule{0.363pt}{0.400pt}}
\multiput(892.00,428.92)(0.839,-0.494){27}{\rule{0.767pt}{0.119pt}}
\multiput(892.00,429.17)(23.409,-15.000){2}{\rule{0.383pt}{0.400pt}}
\multiput(917.00,413.92)(0.901,-0.494){25}{\rule{0.814pt}{0.119pt}}
\multiput(917.00,414.17)(23.310,-14.000){2}{\rule{0.407pt}{0.400pt}}
\multiput(942.00,399.92)(0.901,-0.494){25}{\rule{0.814pt}{0.119pt}}
\multiput(942.00,400.17)(23.310,-14.000){2}{\rule{0.407pt}{0.400pt}}
\multiput(967.00,385.92)(0.972,-0.493){23}{\rule{0.869pt}{0.119pt}}
\multiput(967.00,386.17)(23.196,-13.000){2}{\rule{0.435pt}{0.400pt}}
\multiput(992.00,372.92)(0.933,-0.493){23}{\rule{0.838pt}{0.119pt}}
\multiput(992.00,373.17)(22.260,-13.000){2}{\rule{0.419pt}{0.400pt}}
\multiput(1016.00,359.92)(1.056,-0.492){21}{\rule{0.933pt}{0.119pt}}
\multiput(1016.00,360.17)(23.063,-12.000){2}{\rule{0.467pt}{0.400pt}}
\multiput(1041.00,347.92)(1.156,-0.492){19}{\rule{1.009pt}{0.118pt}}
\multiput(1041.00,348.17)(22.906,-11.000){2}{\rule{0.505pt}{0.400pt}}
\multiput(1066.00,336.92)(1.156,-0.492){19}{\rule{1.009pt}{0.118pt}}
\multiput(1066.00,337.17)(22.906,-11.000){2}{\rule{0.505pt}{0.400pt}}
\multiput(1091.00,325.92)(1.156,-0.492){19}{\rule{1.009pt}{0.118pt}}
\multiput(1091.00,326.17)(22.906,-11.000){2}{\rule{0.505pt}{0.400pt}}
\multiput(1116.00,314.92)(1.277,-0.491){17}{\rule{1.100pt}{0.118pt}}
\multiput(1116.00,315.17)(22.717,-10.000){2}{\rule{0.550pt}{0.400pt}}
\multiput(1141.00,304.93)(1.427,-0.489){15}{\rule{1.211pt}{0.118pt}}
\multiput(1141.00,305.17)(22.486,-9.000){2}{\rule{0.606pt}{0.400pt}}
\multiput(1166.00,295.92)(1.225,-0.491){17}{\rule{1.060pt}{0.118pt}}
\multiput(1166.00,296.17)(21.800,-10.000){2}{\rule{0.530pt}{0.400pt}}
\multiput(1190.00,285.93)(1.616,-0.488){13}{\rule{1.350pt}{0.117pt}}
\multiput(1190.00,286.17)(22.198,-8.000){2}{\rule{0.675pt}{0.400pt}}
\multiput(1215.00,277.93)(1.427,-0.489){15}{\rule{1.211pt}{0.118pt}}
\multiput(1215.00,278.17)(22.486,-9.000){2}{\rule{0.606pt}{0.400pt}}
\multiput(1240.00,268.93)(1.616,-0.488){13}{\rule{1.350pt}{0.117pt}}
\multiput(1240.00,269.17)(22.198,-8.000){2}{\rule{0.675pt}{0.400pt}}
\multiput(1265.00,260.93)(1.616,-0.488){13}{\rule{1.350pt}{0.117pt}}
\multiput(1265.00,261.17)(22.198,-8.000){2}{\rule{0.675pt}{0.400pt}}
\multiput(1290.00,252.93)(1.865,-0.485){11}{\rule{1.529pt}{0.117pt}}
\multiput(1290.00,253.17)(21.827,-7.000){2}{\rule{0.764pt}{0.400pt}}
\multiput(1315.00,245.93)(1.616,-0.488){13}{\rule{1.350pt}{0.117pt}}
\multiput(1315.00,246.17)(22.198,-8.000){2}{\rule{0.675pt}{0.400pt}}
\multiput(1340.00,237.93)(2.118,-0.482){9}{\rule{1.700pt}{0.116pt}}
\multiput(1340.00,238.17)(20.472,-6.000){2}{\rule{0.850pt}{0.400pt}}
\multiput(1364.00,231.93)(1.865,-0.485){11}{\rule{1.529pt}{0.117pt}}
\multiput(1364.00,232.17)(21.827,-7.000){2}{\rule{0.764pt}{0.400pt}}
\multiput(1389.00,224.93)(1.865,-0.485){11}{\rule{1.529pt}{0.117pt}}
\multiput(1389.00,225.17)(21.827,-7.000){2}{\rule{0.764pt}{0.400pt}}
\multiput(1414.00,217.93)(2.208,-0.482){9}{\rule{1.767pt}{0.116pt}}
\multiput(1414.00,218.17)(21.333,-6.000){2}{\rule{0.883pt}{0.400pt}}
\sbox{\plotpoint}{\rule[-0.500pt]{1.000pt}{1.000pt}}%
\put(221,138){\usebox{\plotpoint}}
\multiput(221,138)(1.164,20.723){22}{\usebox{\plotpoint}}
\multiput(246,583)(3.347,20.484){7}{\usebox{\plotpoint}}
\multiput(271,736)(9.911,18.236){3}{\usebox{\plotpoint}}
\put(306.65,785.11){\usebox{\plotpoint}}
\put(326.52,786.91){\usebox{\plotpoint}}
\multiput(345,781)(18.415,-9.576){2}{\usebox{\plotpoint}}
\put(382.59,760.44){\usebox{\plotpoint}}
\multiput(395,753)(17.798,-10.679){2}{\usebox{\plotpoint}}
\put(435.99,728.41){\usebox{\plotpoint}}
\put(453.63,717.48){\usebox{\plotpoint}}
\multiput(470,707)(17.601,-11.000){2}{\usebox{\plotpoint}}
\put(506.46,684.53){\usebox{\plotpoint}}
\multiput(519,677)(17.798,-10.679){2}{\usebox{\plotpoint}}
\put(559.85,652.49){\usebox{\plotpoint}}
\put(577.80,642.07){\usebox{\plotpoint}}
\multiput(594,633)(17.798,-10.679){2}{\usebox{\plotpoint}}
\put(631.69,610.89){\usebox{\plotpoint}}
\multiput(644,604)(17.928,-10.458){2}{\usebox{\plotpoint}}
\put(685.77,580.05){\usebox{\plotpoint}}
\put(703.88,569.91){\usebox{\plotpoint}}
\multiput(718,562)(18.109,-10.141){2}{\usebox{\plotpoint}}
\put(758.21,539.48){\usebox{\plotpoint}}
\put(776.46,529.60){\usebox{\plotpoint}}
\multiput(793,521)(18.109,-10.141){2}{\usebox{\plotpoint}}
\put(831.16,499.87){\usebox{\plotpoint}}
\put(849.48,490.11){\usebox{\plotpoint}}
\multiput(867,481)(18.415,-9.576){2}{\usebox{\plotpoint}}
\put(904.73,461.38){\usebox{\plotpoint}}
\multiput(917,455)(18.712,-8.982){2}{\usebox{\plotpoint}}
\put(960.37,433.45){\usebox{\plotpoint}}
\put(978.78,423.87){\usebox{\plotpoint}}
\multiput(992,417)(18.564,-9.282){2}{\usebox{\plotpoint}}
\put(1034.51,396.11){\usebox{\plotpoint}}
\put(1053.22,387.13){\usebox{\plotpoint}}
\multiput(1066,381)(18.712,-8.982){2}{\usebox{\plotpoint}}
\put(1109.36,360.19){\usebox{\plotpoint}}
\put(1128.07,351.21){\usebox{\plotpoint}}
\multiput(1141,345)(18.998,-8.359){2}{\usebox{\plotpoint}}
\put(1184.43,324.78){\usebox{\plotpoint}}
\put(1203.30,316.15){\usebox{\plotpoint}}
\put(1222.19,307.55){\usebox{\plotpoint}}
\multiput(1240,299)(18.998,-8.359){2}{\usebox{\plotpoint}}
\put(1278.91,281.88){\usebox{\plotpoint}}
\put(1297.91,273.52){\usebox{\plotpoint}}
\multiput(1315,266)(18.998,-8.359){2}{\usebox{\plotpoint}}
\put(1355.03,248.74){\usebox{\plotpoint}}
\put(1374.10,240.56){\usebox{\plotpoint}}
\multiput(1389,234)(19.271,-7.708){2}{\usebox{\plotpoint}}
\put(1431.45,216.32){\usebox{\plotpoint}}
\put(1439,213){\usebox{\plotpoint}}
\put(221,138){\usebox{\plotpoint}}
\multiput(221,138)(1.121,20.725){23}{\usebox{\plotpoint}}
\multiput(246,600)(3.224,20.504){8}{\usebox{\plotpoint}}
\multiput(271,759)(9.747,18.324){2}{\usebox{\plotpoint}}
\put(303.74,808.26){\usebox{\plotpoint}}
\multiput(320,813)(19.529,-7.030){2}{\usebox{\plotpoint}}
\put(361.37,794.83){\usebox{\plotpoint}}
\put(379.15,784.14){\usebox{\plotpoint}}
\multiput(395,774)(17.163,-11.671){2}{\usebox{\plotpoint}}
\put(431.13,749.88){\usebox{\plotpoint}}
\multiput(445,741)(17.163,-11.671){2}{\usebox{\plotpoint}}
\put(482.96,715.36){\usebox{\plotpoint}}
\multiput(494,708)(17.163,-11.671){2}{\usebox{\plotpoint}}
\put(534.80,680.89){\usebox{\plotpoint}}
\put(552.28,669.70){\usebox{\plotpoint}}
\multiput(569,659)(17.482,-11.188){2}{\usebox{\plotpoint}}
\put(604.92,636.45){\usebox{\plotpoint}}
\multiput(619,628)(17.482,-11.188){2}{\usebox{\plotpoint}}
\put(657.71,603.43){\usebox{\plotpoint}}
\put(675.39,592.56){\usebox{\plotpoint}}
\multiput(693,582)(17.798,-10.679){2}{\usebox{\plotpoint}}
\put(728.78,560.53){\usebox{\plotpoint}}
\multiput(743,552)(17.798,-10.679){2}{\usebox{\plotpoint}}
\put(782.18,528.49){\usebox{\plotpoint}}
\put(800.10,518.03){\usebox{\plotpoint}}
\multiput(818,508)(17.601,-11.000){2}{\usebox{\plotpoint}}
\put(853.73,486.43){\usebox{\plotpoint}}
\multiput(867,479)(18.109,-10.141){2}{\usebox{\plotpoint}}
\put(908.06,456.01){\usebox{\plotpoint}}
\put(926.32,446.15){\usebox{\plotpoint}}
\multiput(942,438)(18.109,-10.141){2}{\usebox{\plotpoint}}
\put(980.91,416.21){\usebox{\plotpoint}}
\put(999.08,406.17){\usebox{\plotpoint}}
\multiput(1016,397)(18.415,-9.576){2}{\usebox{\plotpoint}}
\put(1054.17,377.15){\usebox{\plotpoint}}
\multiput(1066,371)(18.415,-9.576){2}{\usebox{\plotpoint}}
\put(1109.41,348.43){\usebox{\plotpoint}}
\put(1127.83,338.85){\usebox{\plotpoint}}
\multiput(1141,332)(18.712,-8.982){2}{\usebox{\plotpoint}}
\put(1183.31,310.62){\usebox{\plotpoint}}
\put(1201.85,301.31){\usebox{\plotpoint}}
\multiput(1215,295)(18.712,-8.982){2}{\usebox{\plotpoint}}
\put(1257.99,274.37){\usebox{\plotpoint}}
\put(1276.70,265.38){\usebox{\plotpoint}}
\multiput(1290,259)(18.712,-8.982){2}{\usebox{\plotpoint}}
\put(1332.83,238.44){\usebox{\plotpoint}}
\put(1351.45,229.27){\usebox{\plotpoint}}
\put(1370.16,220.29){\usebox{\plotpoint}}
\multiput(1389,212)(18.998,-8.359){2}{\usebox{\plotpoint}}
\put(1426.95,194.78){\usebox{\plotpoint}}
\put(1439,189){\usebox{\plotpoint}}
\put(221,140){\usebox{\plotpoint}}
\multiput(221,140)(1.082,20.727){24}{\usebox{\plotpoint}}
\multiput(246,619)(3.128,20.518){8}{\usebox{\plotpoint}}
\multiput(271,783)(9.588,18.408){2}{\usebox{\plotpoint}}
\put(301.84,832.46){\usebox{\plotpoint}}
\multiput(320,837)(19.529,-7.030){2}{\usebox{\plotpoint}}
\put(359.30,818.85){\usebox{\plotpoint}}
\multiput(370,812)(17.163,-11.671){2}{\usebox{\plotpoint}}
\put(410.69,783.71){\usebox{\plotpoint}}
\multiput(420,777)(16.844,-12.128){2}{\usebox{\plotpoint}}
\put(461.22,747.32){\usebox{\plotpoint}}
\put(478.11,735.26){\usebox{\plotpoint}}
\multiput(494,724)(16.844,-12.128){2}{\usebox{\plotpoint}}
\put(528.91,699.26){\usebox{\plotpoint}}
\multiput(544,689)(16.844,-12.128){2}{\usebox{\plotpoint}}
\put(579.93,663.57){\usebox{\plotpoint}}
\multiput(594,654)(17.163,-11.671){2}{\usebox{\plotpoint}}
\put(631.65,628.91){\usebox{\plotpoint}}
\multiput(644,621)(16.937,-11.997){2}{\usebox{\plotpoint}}
\put(683.32,594.20){\usebox{\plotpoint}}
\multiput(693,588)(17.163,-11.671){2}{\usebox{\plotpoint}}
\put(735.30,559.93){\usebox{\plotpoint}}
\put(752.78,548.74){\usebox{\plotpoint}}
\multiput(768,539)(17.798,-10.679){2}{\usebox{\plotpoint}}
\put(805.67,515.89){\usebox{\plotpoint}}
\multiput(818,508)(17.601,-11.000){2}{\usebox{\plotpoint}}
\put(858.28,482.58){\usebox{\plotpoint}}
\put(875.92,471.65){\usebox{\plotpoint}}
\multiput(892,462)(17.798,-10.679){2}{\usebox{\plotpoint}}
\put(929.31,439.61){\usebox{\plotpoint}}
\multiput(942,432)(18.109,-10.141){2}{\usebox{\plotpoint}}
\put(983.14,408.32){\usebox{\plotpoint}}
\put(1001.00,397.75){\usebox{\plotpoint}}
\multiput(1016,389)(17.798,-10.679){2}{\usebox{\plotpoint}}
\put(1054.74,366.31){\usebox{\plotpoint}}
\multiput(1066,360)(18.109,-10.141){2}{\usebox{\plotpoint}}
\put(1109.37,336.45){\usebox{\plotpoint}}
\put(1127.59,326.51){\usebox{\plotpoint}}
\multiput(1141,319)(18.109,-10.141){2}{\usebox{\plotpoint}}
\put(1182.04,296.31){\usebox{\plotpoint}}
\put(1200.38,286.60){\usebox{\plotpoint}}
\multiput(1215,279)(18.109,-10.141){2}{\usebox{\plotpoint}}
\put(1255.21,257.09){\usebox{\plotpoint}}
\put(1273.76,247.80){\usebox{\plotpoint}}
\multiput(1290,240)(18.415,-9.576){2}{\usebox{\plotpoint}}
\put(1329.26,219.58){\usebox{\plotpoint}}
\put(1347.74,210.13){\usebox{\plotpoint}}
\multiput(1364,202)(18.415,-9.576){2}{\usebox{\plotpoint}}
\put(1403.34,182.12){\usebox{\plotpoint}}
\put(1422.05,173.13){\usebox{\plotpoint}}
\put(1439,165){\usebox{\plotpoint}}
\put(221,165){\usebox{\plotpoint}}
\multiput(221,165)(1.151,20.724){22}{\usebox{\plotpoint}}
\multiput(246,615)(3.326,20.487){8}{\usebox{\plotpoint}}
\multiput(271,769)(10.080,18.144){2}{\usebox{\plotpoint}}
\put(301.81,815.45){\usebox{\plotpoint}}
\multiput(320,820)(19.271,-7.708){2}{\usebox{\plotpoint}}
\put(359.47,801.89){\usebox{\plotpoint}}
\multiput(370,796)(17.163,-11.671){2}{\usebox{\plotpoint}}
\put(411.51,767.77){\usebox{\plotpoint}}
\put(428.68,756.10){\usebox{\plotpoint}}
\multiput(445,745)(17.163,-11.671){2}{\usebox{\plotpoint}}
\put(480.03,720.89){\usebox{\plotpoint}}
\multiput(494,711)(17.482,-11.188){2}{\usebox{\plotpoint}}
\put(531.79,686.30){\usebox{\plotpoint}}
\multiput(544,678)(17.482,-11.188){2}{\usebox{\plotpoint}}
\put(584.01,652.39){\usebox{\plotpoint}}
\multiput(594,646)(17.482,-11.188){2}{\usebox{\plotpoint}}
\put(636.46,618.83){\usebox{\plotpoint}}
\put(653.82,607.45){\usebox{\plotpoint}}
\multiput(668,598)(17.798,-10.679){2}{\usebox{\plotpoint}}
\put(706.53,574.34){\usebox{\plotpoint}}
\multiput(718,567)(17.798,-10.679){2}{\usebox{\plotpoint}}
\put(759.72,541.97){\usebox{\plotpoint}}
\put(777.52,531.29){\usebox{\plotpoint}}
\multiput(793,522)(17.798,-10.679){2}{\usebox{\plotpoint}}
\put(830.77,499.02){\usebox{\plotpoint}}
\multiput(842,492)(18.109,-10.141){2}{\usebox{\plotpoint}}
\put(884.46,467.52){\usebox{\plotpoint}}
\put(902.44,457.15){\usebox{\plotpoint}}
\multiput(917,449)(18.109,-10.141){2}{\usebox{\plotpoint}}
\put(956.77,426.73){\usebox{\plotpoint}}
\put(974.88,416.59){\usebox{\plotpoint}}
\multiput(992,407)(18.250,-9.885){2}{\usebox{\plotpoint}}
\put(1029.39,386.50){\usebox{\plotpoint}}
\put(1047.61,376.56){\usebox{\plotpoint}}
\multiput(1066,367)(18.415,-9.576){2}{\usebox{\plotpoint}}
\put(1102.66,347.47){\usebox{\plotpoint}}
\multiput(1116,340)(18.415,-9.576){2}{\usebox{\plotpoint}}
\put(1157.95,318.87){\usebox{\plotpoint}}
\put(1176.39,309.37){\usebox{\plotpoint}}
\multiput(1190,302)(18.415,-9.576){2}{\usebox{\plotpoint}}
\put(1231.78,280.94){\usebox{\plotpoint}}
\put(1250.33,271.63){\usebox{\plotpoint}}
\multiput(1265,264)(18.712,-8.982){2}{\usebox{\plotpoint}}
\put(1306.22,244.21){\usebox{\plotpoint}}
\put(1324.94,235.23){\usebox{\plotpoint}}
\multiput(1340,228)(18.564,-9.282){2}{\usebox{\plotpoint}}
\put(1380.88,207.90){\usebox{\plotpoint}}
\put(1399.75,199.27){\usebox{\plotpoint}}
\multiput(1414,193)(18.712,-8.982){2}{\usebox{\plotpoint}}
\put(1439,181){\usebox{\plotpoint}}
\put(221,152){\usebox{\plotpoint}}
\multiput(221,152)(1.131,20.725){23}{\usebox{\plotpoint}}
\multiput(246,610)(3.264,20.497){7}{\usebox{\plotpoint}}
\multiput(271,767)(9.911,18.236){3}{\usebox{\plotpoint}}
\put(310.47,816.62){\usebox{\plotpoint}}
\put(330.15,814.94){\usebox{\plotpoint}}
\multiput(345,809)(18.109,-10.141){2}{\usebox{\plotpoint}}
\put(384.84,785.50){\usebox{\plotpoint}}
\multiput(395,779)(17.163,-11.671){2}{\usebox{\plotpoint}}
\put(436.52,750.77){\usebox{\plotpoint}}
\put(453.68,739.10){\usebox{\plotpoint}}
\multiput(470,728)(16.937,-11.997){2}{\usebox{\plotpoint}}
\put(505.05,703.93){\usebox{\plotpoint}}
\multiput(519,695)(17.163,-11.671){2}{\usebox{\plotpoint}}
\put(557.03,669.66){\usebox{\plotpoint}}
\multiput(569,662)(17.482,-11.188){2}{\usebox{\plotpoint}}
\put(609.48,636.09){\usebox{\plotpoint}}
\put(626.96,624.91){\usebox{\plotpoint}}
\multiput(644,614)(17.270,-11.513){2}{\usebox{\plotpoint}}
\put(679.31,591.21){\usebox{\plotpoint}}
\multiput(693,583)(17.482,-11.188){2}{\usebox{\plotpoint}}
\put(732.25,558.45){\usebox{\plotpoint}}
\multiput(743,552)(17.798,-10.679){2}{\usebox{\plotpoint}}
\put(785.64,526.41){\usebox{\plotpoint}}
\put(803.44,515.73){\usebox{\plotpoint}}
\multiput(818,507)(17.928,-10.458){2}{\usebox{\plotpoint}}
\put(857.01,483.99){\usebox{\plotpoint}}
\put(874.94,473.55){\usebox{\plotpoint}}
\multiput(892,464)(18.109,-10.141){2}{\usebox{\plotpoint}}
\put(929.27,443.13){\usebox{\plotpoint}}
\multiput(942,436)(18.109,-10.141){2}{\usebox{\plotpoint}}
\put(983.60,412.70){\usebox{\plotpoint}}
\put(1001.79,402.70){\usebox{\plotpoint}}
\multiput(1016,395)(18.109,-10.141){2}{\usebox{\plotpoint}}
\put(1056.48,372.95){\usebox{\plotpoint}}
\put(1074.89,363.38){\usebox{\plotpoint}}
\multiput(1091,355)(18.109,-10.141){2}{\usebox{\plotpoint}}
\put(1129.72,333.87){\usebox{\plotpoint}}
\put(1148.25,324.52){\usebox{\plotpoint}}
\multiput(1166,316)(18.250,-9.885){2}{\usebox{\plotpoint}}
\put(1203.56,295.95){\usebox{\plotpoint}}
\put(1222.08,286.60){\usebox{\plotpoint}}
\multiput(1240,278)(18.712,-8.982){2}{\usebox{\plotpoint}}
\put(1278.01,259.24){\usebox{\plotpoint}}
\put(1296.53,249.87){\usebox{\plotpoint}}
\multiput(1315,241)(18.712,-8.982){2}{\usebox{\plotpoint}}
\put(1352.77,223.15){\usebox{\plotpoint}}
\put(1371.57,214.37){\usebox{\plotpoint}}
\multiput(1389,206)(18.712,-8.982){2}{\usebox{\plotpoint}}
\put(1427.92,187.88){\usebox{\plotpoint}}
\put(1439,183){\usebox{\plotpoint}}
\end{picture}